\documentclass[final,3p,sort&compress,times,twocolumn,10pt]{elsarticle}
\usepackage[utf8]{inputenc}
\usepackage{color}
\usepackage{amsfonts,amsmath,amssymb}
\usepackage{longtable,booktabs}
\usepackage{graphicx}
\usepackage{bm}
\usepackage[colorlinks=true,linkcolor=blue,citecolor=blue]{hyperref}
\usepackage{epstopdf}
\usepackage{mathtools}
\usepackage{subfigure}
\usepackage{multirow}
\usepackage{diagbox}
\usepackage{afterpage}
\usepackage{placeins}
\usepackage{float}
\usepackage{mathrsfs}
\usepackage{tabularray}
\biboptions{sort&compress}

\journal{Physics Letters B}

\newcolumntype{M}[1]{>{\centering\arraybackslash}m{#1}}
\newcolumntype{N}{@{}m{0pt}@{}}

\begin{document}
\begin{frontmatter}
\title{Classification of Coupled-Channel Near-Threshold Structures}

\author[chep,itp]{Zhen-Hua Zhang}\ead{zhhzhang@pku.edu.cn}

\author[itp,ucas,peng,scnt]{Feng-Kun Guo\corref{corresponding}}\ead{fkguo@itp.ac.cn} 

\address[chep]{Center for High Energy Physics, Peking University, Beijing 100871, China}
\address[itp]{CAS Key Laboratory of Theoretical Physics, Institute of Theoretical Physics,\\
	Chinese Academy of Sciences, Beijing 100190, China}
\address[ucas]{School of Physical Sciences, University of Chinese Academy of Sciences, Beijing 100049, China}
\address[peng]{Peng Huanwu Collaborative Center for Research and Education, Beihang University, Beijing 100191, China}
\address[scnt]{Southern Center for Nuclear-Science Theory (SCNT), Institute of Modern Physics,\\ Chinese Academy of Sciences, Huizhou 516000, China}

\cortext[corresponding]{Corresponding author.}
\begin{abstract}

Since 2003, plenty of resonant structures have been observed in the heavy quarkonium regime. Many of them are close to the thresholds of a few pairs of heavy hadrons.
They are candidates of exotic hadrons and have attracted immense attentions.
Based on a coupled-channel nonrelativistic effective field theory, we classify the near-threshold structures of a symmetry-related two-channel system by studying the evolution of the scattering amplitude line shapes and pole positions with the variation of the single-channel scattering length and channel coupling strength.
We show that the evolution of the scattering amplitude line shapes can be understood from the pole trajectories in the complex energy plane, and the pole evolution can be traced back to the renormalization group fixed points.
We provide a dictionary of correspondence between the evolution of line shapes and pole trajectories along with  varying interaction and channel coupling strengths, which can be used to understand the experimental observations of the near-threshold structures.

\end{abstract}

\end{frontmatter}



\section{Introduction}

In the past two decades, plethora of resonant structures have been observed in the invariant mass 
distribution of heavy hadrons at the high energy experiment, and have attracted lots of attentions and debates.
Strenuous efforts have been made to understand how these resonances emerge from underlying strong interactions and what internal structures they have, 
while consensus on these questions has not been achieved (for reviews, see Refs.~\cite{Esposito:2016noz, Hosaka:2016pey, Guo:2017jvc, Olsen:2017bmm, Karliner:2017qhf, Brambilla:2019esw, Guo:2019twa, Chen:2022asf, Meng:2022ozq}). 

A surprising and prominent feature is that many of the resonant structures are close to the threshold of a pair of hadrons containing heavy quarks,
{\it e.g.}, the famous $X(3872)$ is located in the immediate vicinity of the $D^{0}\bar{D}^{*0}$ threshold~\cite{Belle:2003nnu,LHCb:2020xds,LHCb:2020fvo}. 
A general explanation about these near-threshold structures has been given in Ref.~\cite{Dong:2020hxe}, which shows that a nontrivial peak or dip structure close to the threshold of a pair of heavy hadrons must appear if they have $S$-wave attractive interaction.
However, such construction still needs to be generalized to situations where the resonant structures are close to more than one threshold, which is often the physical situation and the line shapes can be more complicated because of the intertwined energy dependence of the amplitude caused by multiple thresholds.
For instance, among recent experimental discoveries, while the $X(3872)$ mass coincides with the $D^0\bar D^{*0}$ threshold within uncertainties~\cite{ParticleDataGroup:2024}, the $D^+D^{*-}$ threshold is only about 8~MeV away; the $X(6900)$ in 
the double $J/\psi$ spectrum~\cite{LHCb:2020bwg, ATLAS:2023bft, CMS:2023owd} is close to the $J/\psi\psi(2S)$ and $J/\psi\psi(3770)$ thresholds~\cite{Dong:2020nwy}; 
the $T_{c\bar{s}}(2900)$ seen in the $D_s^+\pi^-$ and $D_s^+\pi^+$ invariant mass distributions
in the $B^0\to \bar{D}^0 D_s^+\pi^-$ and $B^+\to D^-D_s^+\pi^+$ decays~\cite{LHCb:2022sfr, LHCb:2022lzp} is close to the $D_s^*\rho$ and $D^*K^*$ thresholds~\cite{Molina:2022jcd};
the $T_{cc}^+(3875)$ in the $D^0D^0\pi^{+}$ invariant mass distribution~\cite{LHCb:2021vvq,LHCb:2021auc} is just below the $D^{*+}D^0$ and 
$D^{*0}D^+$ thresholds~\cite{Fleming:2021wmk,Albaladejo:2021vln,Du:2021zzh}; the $P_{cs}(4338)^0$ observed in 
the $J/\psi\Lambda$ spectrum~\cite{LHCb:2022ogu} is very close to the $\Xi_c^0\bar{D}^0$ and $\Xi_c^+D^-$ thresholds~\cite{Meng:2022wgl};  
and the $X(3960)$ in the $D_s^+D_s^-$ spectrum~\cite{LHCb:2022aki,LHCb:2022dvn} is close to the $D_s^+D_s^-$ and $D^{*}\bar{D}^{*}$ thresholds~\cite{Ji:2022uie}; and so on.
These are just some representatives of the structures close to at least two nearby thresholds, and most of 
the nearby channels are related by some kind of symmetry, {\it e.g.}, isospin symmetry, SU(3) symmetry, and/or heavy quark spin symmetry.
The resonance line shapes are intricately distorted by these thresholds, and a classification of the general behavior near multiple thresholds is called for. 

In this Letter, we classify the general line shape behavior near multiple thresholds with variations of the single-channel scattering length and channel coupling strength, starting from the renormalization group (RG) fixed points (FPs) in the framework of a coupled-channel zero-range effective field theory (ZREFT). 
For simplicity, we will discuss a system with two symmetry-related nearby channels and the discussion can be generalized to systems with more channels.

\section{Renormalization group fixed points}

We consider a two-channel system (channel-1 and channel-2) with nearby thresholds denoted as $\Sigma_1$ and $\Sigma_2$, with $\Delta\equiv\Sigma_2-\Sigma_1>0$. The corresponding reduced masses are represented by $M_1$ and $M_2$, respectively. We will focus on the line shapes near the 
two thresholds, and therefore both channels will be treated nonrelativistically. We define the center-of-mass (c.m.) energy relative to 
the first threshold as $E=\sqrt{s}-\Sigma_1$, with $\sqrt{s}$ being the total c.m. energy.

We first review the RG treatment to the two-channel scattering derived in Ref.~\cite{Lensky:2011he}.
For the on-shell $S$-wave scattering processes, the scattering potential only depends on the c.m. energy~\cite{Lensky:2011he} ({\it i.e.}, separable potential),
and thus the Lippmann-Schwinger equation (LSE) for the two-channel scattering amplitude can be 
written as an algebraic equation
\begin{equation}
    \mathbf{T}(p_1,\delta_1)=\mathbf{V}(p_1,\delta_1)+\mathbf{V}(p_1,\delta_1)\mathbf{J}(p_1,\delta_1)\mathbf{T}(p_1,\delta_1),
    \label{Eq.LSE}
\end{equation}
where $p_1=\sqrt{2M_1E}$ and $\delta_1=\sqrt{2M_1\Delta}$ are small momentum scales, the ultraviolet divergent Green's function $\mathbf{J}(p_1,\delta_1)$ can be regularized by the dimensional regularization with the power divergence subtraction scheme~\cite{Kaplan:1998tg,Kaplan:1998we} as
\begin{align}
    \mathbf{J}(p_1,\delta_1,\mu)=-\frac{1}{2\pi}\mathbf{M}^{1/2}(\mu \mathbf{I}_{2\times2}+i\mathbf{P})\mathbf{M}^{1/2},
\end{align}
where $\mu$ is the subtraction scale, $\mathbf{P}=\mathrm{diag}(p_1,p_2)$ with $p_1$ and $p_2=\sqrt{2M_2(E-\Delta)}$ the magnitudes of the c.m. momenta of particles in
channel-1 and channel-2, respectively, and $\mathbf{M} = \mathrm{diag}(M_1,M_2)$.

The scale independence of the scattering amplitude gives 
the RG equation (RGE) of the potential $\mathbf{V}(p_1,\delta_1,\mu)$,
\begin{align}
    \mu\frac{\partial\mathbf{\hat{V}}}{\partial \mu}=\hat{p}\frac{\partial\mathbf{\hat{V}}}{\partial \hat{p}}+\hat{\delta}\frac{\partial\mathbf{\hat{V}}}{\partial \hat{\delta}}+\mathbf{\hat{V}}+\mathbf{\hat{V}}^2,
    \label{Eq.RGE_standard}
\end{align}
where $\hat{p}\equiv {p_1}/{\mu}$, $\hat{\delta}\equiv {\delta_1}/{\mu}$, and
$\mathbf{\hat{V}}={\mu}/{(2\pi)}\mathbf{M}^{1/2}\mathbf{V}\mathbf{M}^{1/2}$. 
Its scale invariant solutions are the RG FPs. There are three types of FPs for a two-channel system~\cite{Lensky:2011he}: the trivial FP corresponding to vanishing interaction:
\begin{equation}
    \mathbf{\hat{V}}_0=0, \label{eq:1stFP}
\end{equation}
the FP with two bound/virtual states at the threshold:
\begin{equation}
    \mathbf{\hat{V}}_2=-\mathbf{I}_{2\times 2}, \label{eq:2ndFP}
\end{equation}
and the FP $\hat{\mathbf{V}}_1$ with only one bound/virtual state at the threshold:
\begin{align}
    \mathbf{\hat{V}}_1= \begin{pmatrix}
       -c & \pm \sqrt{c(1-c)} \\
       \pm\sqrt{c(1-c)} & - (1-c) 
    \end{pmatrix}. \label{eq:3rdFP}
\end{align}
which is noninvertable, with $c$ a real parameter.

The general scattering amplitude can be derived from the LSE using the potential solved from the RGE. That is, it can be obtained by a power series expansion in $\hat{p}$ and $\hat{\delta}$~\cite{Lensky:2011he} in the vicinity of the RG FP $\hat{\mathbf{V}}_2$.
Notice that with $M_1\sim M_2$, around the thresholds, we have $\delta_1 \sim p_2$.
Keeping only constant contact terms in the potential series, one obtains the leading order (LO) scattering amplitude as~\cite{Lensky:2011he}
\begin{align}
    \mathbf{T}^{\mathrm{LO}}=2\pi\mathbf{M}^{-1/2}\mathbf{R}\begin{pmatrix}
        -\frac{1}{a_{11}}+ip_{11} &\frac{1}{a_{12}}+ip_{12}\\
        \frac{1}{a_{12}}+ip_{12}& -\frac{1}{a_{22}}+ip_{22}
    \end{pmatrix}^{-1}
     \mathbf{R}^{T}\mathbf{M}^{-1/2},
    \label{Eq.Zero_range_T_matrix_two_channels}
\end{align}
with 
\begin{equation}
    \mathbf{R} = \begin{pmatrix}
        \cos{\phi} & -\sin{\phi}\\
        \sin{\phi}&\cos{\phi}        
    \end{pmatrix} \in \operatorname{SO}(2)
    \label{eq:R}
\end{equation}
a rotation between the two channels, $\phi$ the rotation angle and $p_{ij}=\left(\mathbf{R}^{T}\mathbf{P}\mathbf{R}\right)_{ij}$. The parameters $1/a_{ij}$ correspond to the LO expansion coefficients in the potential series. The values of $a_{ij}$ measure the interaction strength in/between the two channels.
Parameters $a_{11}$ and $a_{22}$ are the single-channel scattering lengths.
The channel coupling is provided by a finite $|a_{12}|$ and induces an effective attraction in channel-1 and repulsion in channel-2 (see, {\it e.g.}, Refs.~\cite{Guo:2016bjq, Molina:2022jcd}).
Near the FP $\mathbf{\hat{V}}_2$ in Eq.~\eqref{eq:2ndFP}, both $a_{11}$ and $a_{22}$ are unnaturally large, one has $|a_{11}^{-1}|< \mu$, $|a_{22}^{-1}|< \mu$.\footnote{Here $\mu\gg p_1,\delta_1$ can be understood as a reference scale for choosing different scalings of $a_{ij}$ to perform expansions around the RG FPs.} The LO scattering amplitude in Eq.~\eqref{Eq.Zero_range_T_matrix_two_channels} describes a system with two bound/virtual states near the thresholds~\cite{Cohen:2004kf}.

Near other different FPs, the $a_{ij}$ parameters have different scalings and the LO $\mathbf{T}$ matrix in Eq.~\eqref{Eq.Zero_range_T_matrix_two_channels} can be further expanded and simplified~\cite{Lensky:2011he}. The amplitude near the trivial FP $\mathbf{\hat{V}}_0=0$ describes a weakly interacting system and is not of interest here.

The FP $\hat{\mathbf{V}}_1$ can be expressed by the rotation angle $\phi$ as

\begin{align}
    \mathbf{\hat{V}}_1=-\mathbf{R}\begin{pmatrix}
        1&0\\
        0&0
\end{pmatrix} \mathbf{R}^T
=-\begin{pmatrix}
    \cos^2\phi& \sin\phi\cos\phi\\
    \sin\phi\cos\phi& \sin^2\phi
\end{pmatrix},
\label{eq:3rdFP_2}
\end{align}
which is just Eq.~\eqref{eq:3rdFP} with $c=\cos^2\phi$.
Near the FP $\mathbf{\hat{V}}_1$~\cite{Lensky:2011he}, only one of the single-channel scattering lengths $a_{11}$ and $a_{22}$ is unnaturally large, one can take $|a_{11}^{-1}|< \mu$, $|a_{22}^{-1}|\gg \mu$, and the LO amplitude reads 
\begin{align}
    \mathbf{T}^{\mathrm{LO}}=\frac{2\pi}{-a_{11}^{-1}+ip_{11}}\mathbf{M}^{-1/2} (-\hat{\mathbf{V}}_1) \mathbf{M}^{-1/2}, 
    \label{Eq.T_matrix_one_state}
\end{align}
which describes a system with one near-threshold bound/virtual state.

\section{Line shapes and poles}
\label{sec:lspoles}

\begin{figure}[tb]
    \centering
    \includegraphics[width=0.8\columnwidth]{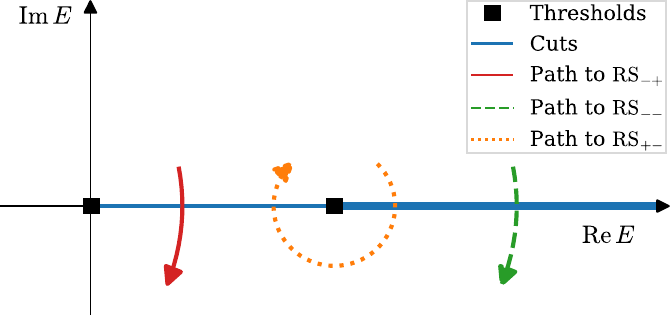}
    \caption{The shortest paths from the physical region to different unphysical RSs.  }
    \label{Fig.RSs}
\end{figure}

With these LO scattering amplitudes, one can classify the LO near-threshold behaviors of the scattering amplitude line shapes according to the pole locations evolved from the RG FPs with the variation of $a_{ij}$. As is well-known (see, {\it e.g.}, Refs.~\cite{Guo:2017jvc,Brambilla:2019esw,Dong:2020hxe}), if the pole is located on a Riemann sheet (RS) that can reach the physical region only by going around a threshold, its effects will manifest as a cusp, exactly at threshold. The width of the cusp depends on the distance of the pole to the physics region.
For simplicity, we consider symmetry related channels by imposing $a_{22}=a_{11}$ in Eq.~\eqref{Eq.Zero_range_T_matrix_two_channels}. 
Without loss of generality, we will show  pole trajectories and line shapes with the masses of the particles in the two channels being those of $D^{0}\bar{D}^{*0}$ and $D^{+}D^{*-}$, respectively. 
These two channels are related by the isospin symmetry and are relevant for the isoscalar $X(3872)$ with $J^{PC}=1^{++}$, the isovector $Z_c(3900)$ with $J^{PC}=1^{+-}$, and the isovector $W_{c1}$ with $J^{PC}=1^{++}$ predicted in the hadronic molecular picture in Ref.~\cite{Zhang:2024fxy}. For this system, $\delta\equiv \sqrt{2M_2\Delta}$ equals 0.64~fm$^{-1}$. 

As a function on the complex energy ($E$) plane with branch cuts $[0,+\infty)$ and  $[\Delta,+\infty)$ along the positive real-$E$ axis, the two-channel scattering amplitudes have four RSs.
The RSs are denoted as $\mathrm{RS}_{r_1 r_2}$ with the subindex representing the signs of $\mathrm{Im}p_1$ and $\mathrm{Im}p_2$;
$\mathrm{RS}_{++}$, $\mathrm{RS}_{-+}$, $\mathrm{RS}_{--}$ and $\mathrm{RS}_{+-}$ correspond to the first to the fourth RS, respectively. 
The first RS is also called the physical RS, whereas the other three are unphysical ones.
The shortest 
path to each unphysical RS from the physical region (the upper edge of the cut in $\mathrm{RS}_{++}$) is shown in Fig.~\ref{Fig.RSs}. 
RS$_{-+}$ and RS$_{--}$ can be reached 
by crossing the cuts only once and thus are directly connected to the physical region, while RS$_{+-}$ can only be reached by crossing the cuts twice.

\begin{figure}[tb]
    \centering
    \includegraphics[width=\linewidth]{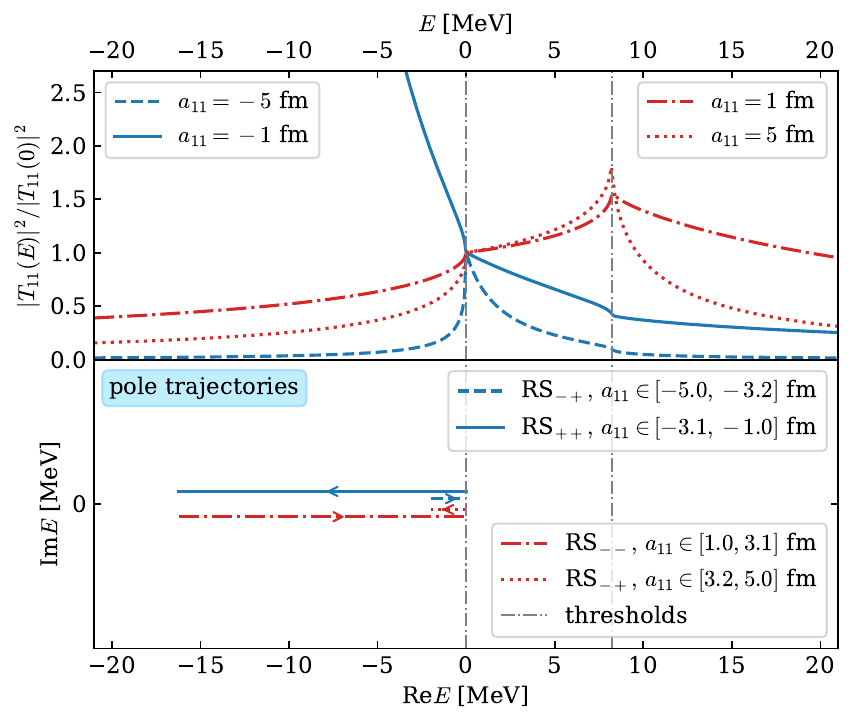}
    \caption{Line shapes (upper panel) and pole trajectories (lower panel) for the one FP case.}
    \label{fig:1fp}
\end{figure}
Let us first consider the evolution from the single-pole FP with the variation of $a_{11}$.
For the two channels related by a symmetry, the diagonal matrix elements of the potential matrix $\hat{\mathbf{V}}$ should be the same. Noninvertability of $\hat{\mathbf{V}}$ for the single-pole FP case then requires it to take the form 
\begin{align}
    \hat{\mathbf{V}}\propto \begin{pmatrix}
        1&\pm 1\\
        \pm 1 & 1
    \end{pmatrix},
\end{align}
and thus we have $c=\cos^2\phi=\sin^2\phi=1/2$ in Eqs.~\eqref{eq:3rdFP} and \eqref{eq:3rdFP_2} as the FP. Evolving $a_{11}$ away from infinity, the pole of Eq.~\eqref{Eq.T_matrix_one_state} is located at the solution of
\begin{align}
\frac{2}{a_{11}}-ir_1p_1-ir_2p_2 = 0, 
\label{eq:pole1c}
\end{align}
on RS$_{r_1 r_2}$. 
Equation~\eqref{eq:pole1c} implies that amplitudes with opposite $a_{11}$ have the same pole positions but on RSs with opposite subindices. 
This will be called pole duality and the corresponding RSs will be called dual RSs in the following.
For $a_{11}\delta< -2$, the effective scattering length $a_{11,\mathrm{eff}}=({2}/{a_{11}}+\delta)^{-1}$ in channel-1 is positive, and one gets a virtual state pole located on RS$_{-+}$. Increasing $a_{11}$ such that $a_{11,\mathrm{eff}}$ becomes negative, the pole moves to RS$_{++}$ and becomes a bound state pole. Correspondingly, the line shape changes from a narrow threshold cusp to a below-threshold peak, as shown as the blue lines in Fig.~\ref{fig:1fp} (line shapes in the upper panel and pole trajectory in the lower panel).
The trajectory of the pole for positive $a_{11}$ on the dual RS and the corresponding line shapes are shown as the red lines in Fig.~\ref{fig:1fp}.
One sees that there is always a single peak in the line shape. This is the case of Fit 2 in the analysis of the $Z_c$ and $Z_{cs}$ states in Ref.~\cite{Baru:2021ddn}, which has more complications due to the existence of triangle singularities~\cite{Guo:2019twa, Wang:2013cya}.

\begin{table*}[tbp]
    \centering
    \caption{Line shapes of $T$-matrix elements in the near-threshold region for the negative $a_{11}$ cases. Pole trajectories evolving among the cases are shown in the last column. The shadow pole of pole-1 is denoted as Spole-1.}
    \begin{tblr}{width=\textwidth,colspec={X[l,1em] X[c,2.8em] X[c,2em] Q[c,m] |Q[c,m]}}
        \hline\hline 
        Case & $a_{11}\delta$ & $|a_{12}|\delta$ & Line shapes & Pole trajectories \\ \hline 
        \SetCell[r=2]{m}B1 & \SetCell[r=2]{m} $\ll -1$ & \SetCell[r=2]{m} $\gg 1$ & 
        \SetCell[r=2]{m,0.38\textwidth} \includegraphics[width=\linewidth]{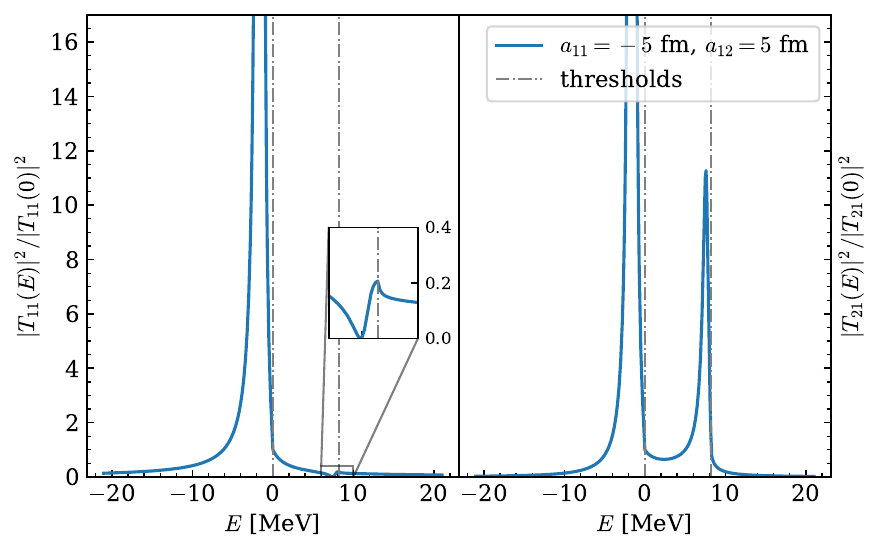}\label{Fig.ls_B1} &  \SetCell[r=2]{m,0.38\textwidth} \includegraphics[width=\linewidth]{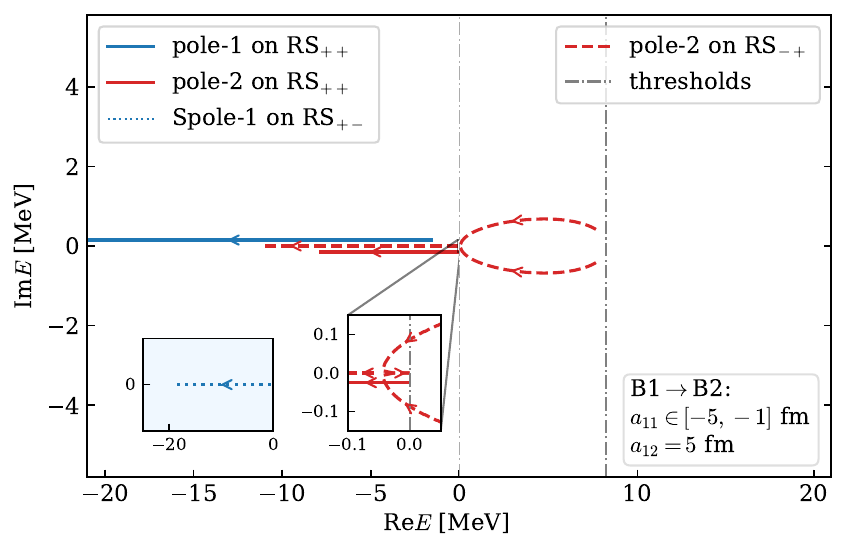} \\ & & & & \\ \cline{1-4}  \\
        \SetCell[r=2]{m}B2 & \SetCell[r=2]{m} $(-1,0)$ & \SetCell[r=2]{m} $\gg 1$ & 
        \SetCell[r=2]{m,0.38\textwidth} \includegraphics[width=\linewidth]{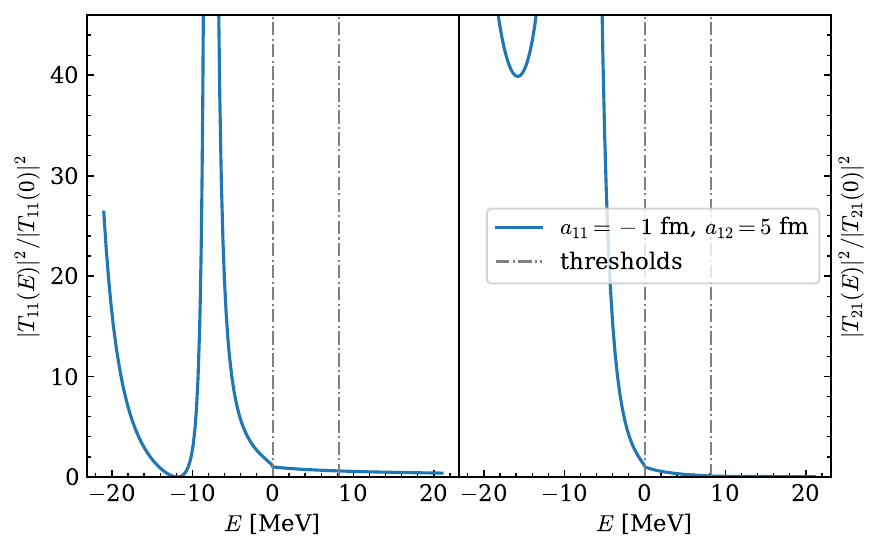} & \SetCell[r=2]{m,0.38\textwidth} \includegraphics[width=\linewidth]{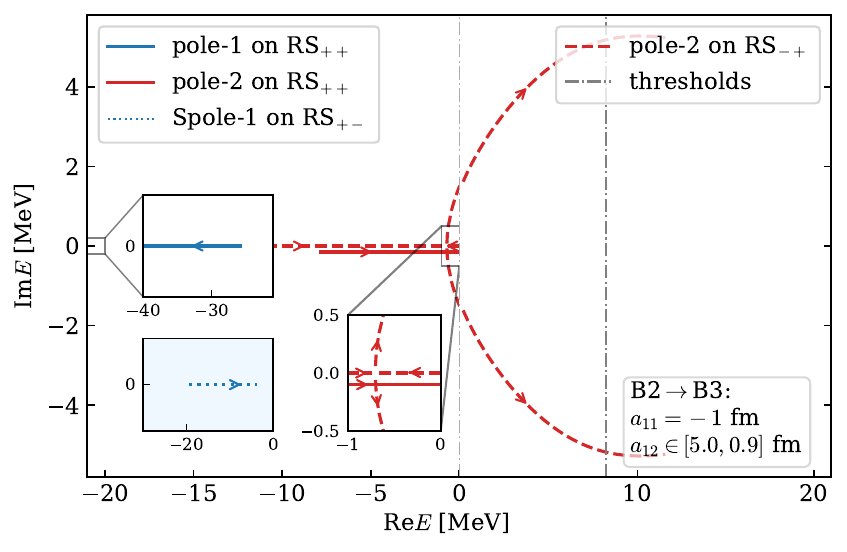}\\  & & & & \\ \cline{1-4} \\
        \SetCell[r=2]{m}B3 & \SetCell[r=2]{m} $(-1,0)$ & \SetCell[r=2]{m} $(0,1)$ &
        \SetCell[r=2]{m,0.4\textwidth} \includegraphics[width=\linewidth]{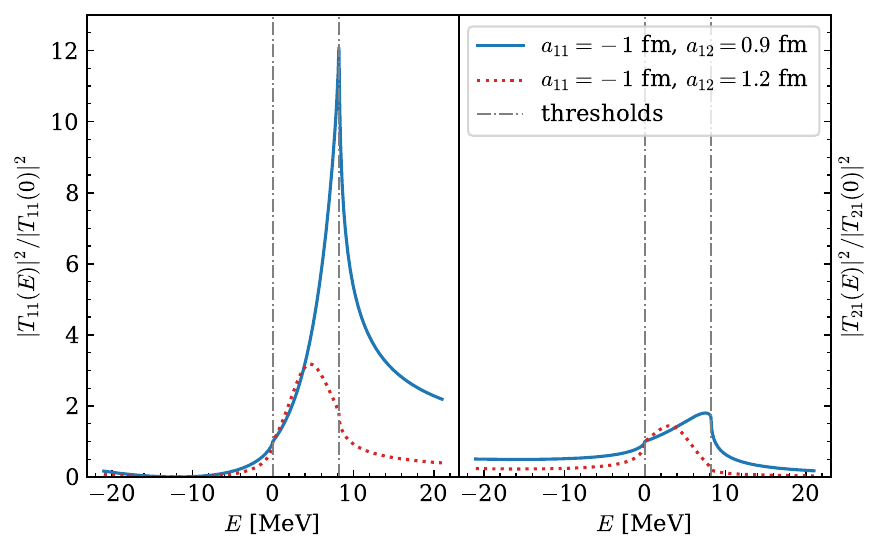} & \SetCell[r=2]{m,0.38\textwidth} \includegraphics[width=\linewidth]{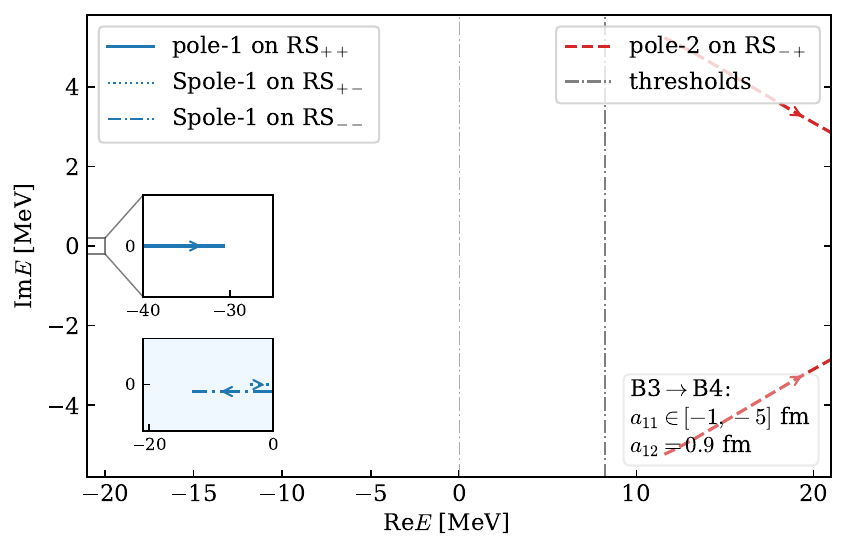}\\  & & & & \\ \cline{1-4} \\
        \SetCell[r=2]{m}B4 & \SetCell[r=2]{m} $\ll -1$ & \SetCell[r=2]{m} $(0,1)$ &
        \SetCell[r=2]{m,0.38\textwidth} \includegraphics[width=\linewidth]{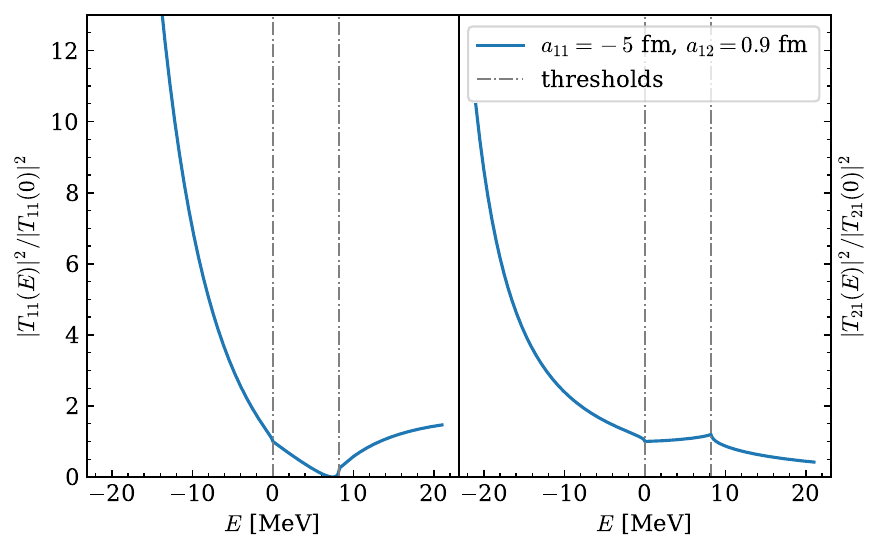} & \SetCell[r=2]{m,0.38\textwidth} \includegraphics[width=\linewidth]{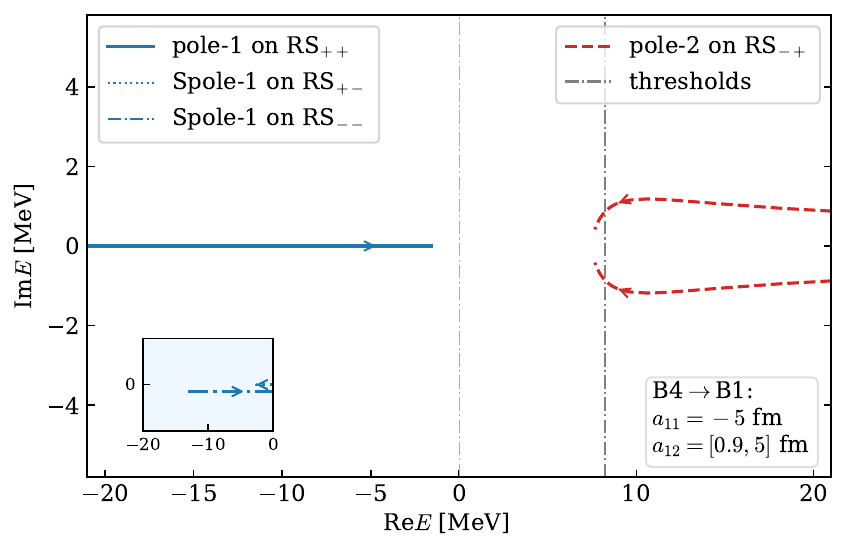}\\ 
        & & & & \\ \hline\hline
    \end{tblr} 
    \label{tab:classification_bs}
\end{table*}

\begin{table*}[tbp]
    \centering
    \caption{Line shapes of $T$-matrix elements in the near-threshold region for the positive $a_{11}$ cases. Pole trajectories evolving among the cases are shown in the last column. The shadow pole of pole-1 is denoted as Spole-1. }
    \begin{tblr}{width=\textwidth,colspec={X[l,1em] X[c,2.8em] X[c,2em] Q[c,m] |Q[c,m]}}
        \hline\hline 
        Case & $a_{11}\delta$ & $|a_{12}|\delta$ & Line shapes & Pole trajectories \\ \hline 
        \SetCell[r=2]{m}V1 & \SetCell[r=2]{m} $\gg 1$ & \SetCell[r=2]{m} $\gg 1$ & 
        \SetCell[r=2]{m,0.38\textwidth} \includegraphics[width=\linewidth]{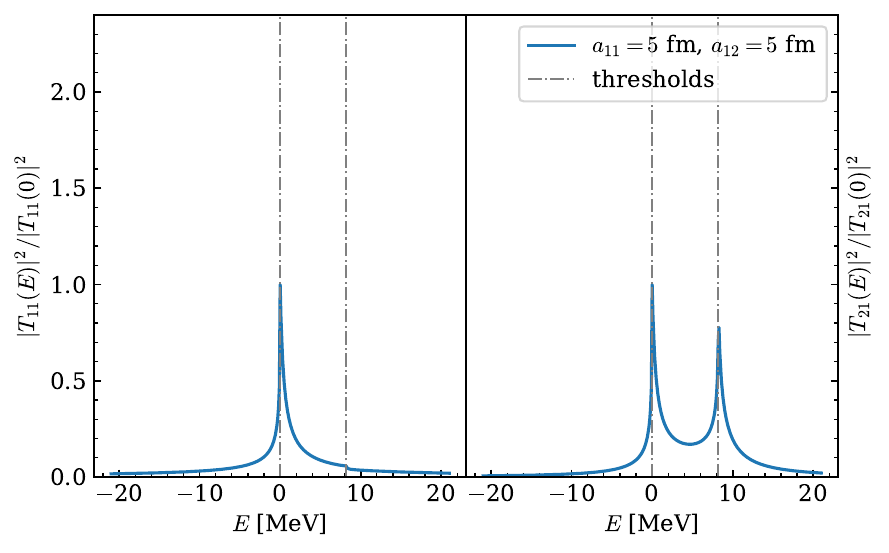} & \SetCell[r=2]{m,0.38\textwidth} \includegraphics[width=\linewidth]{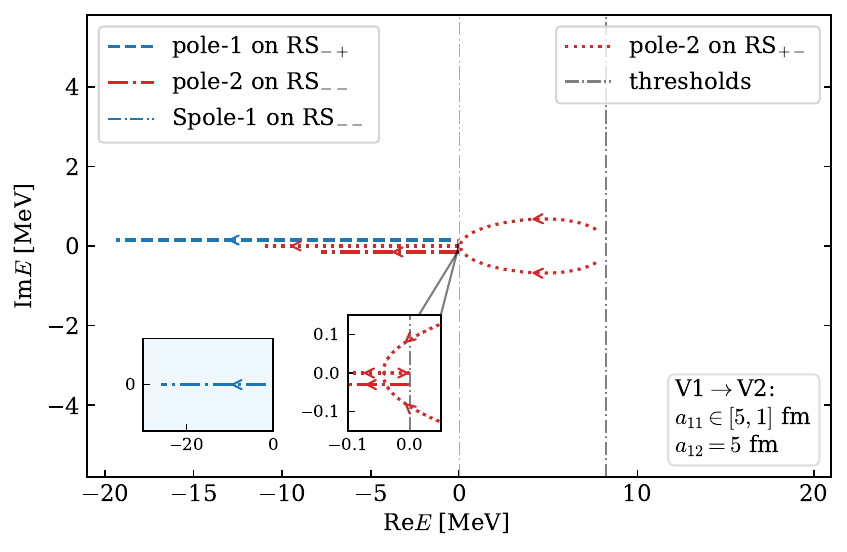}\\ & & & &\\ \cline{1-4}\\ 
        \SetCell[r=2]{m}V2 & \SetCell[r=2]{m} $(0,1)$ & \SetCell[r=2]{m} $\gg 1$ & 
        \SetCell[r=2]{m,0.38\textwidth} \includegraphics[width=\linewidth]{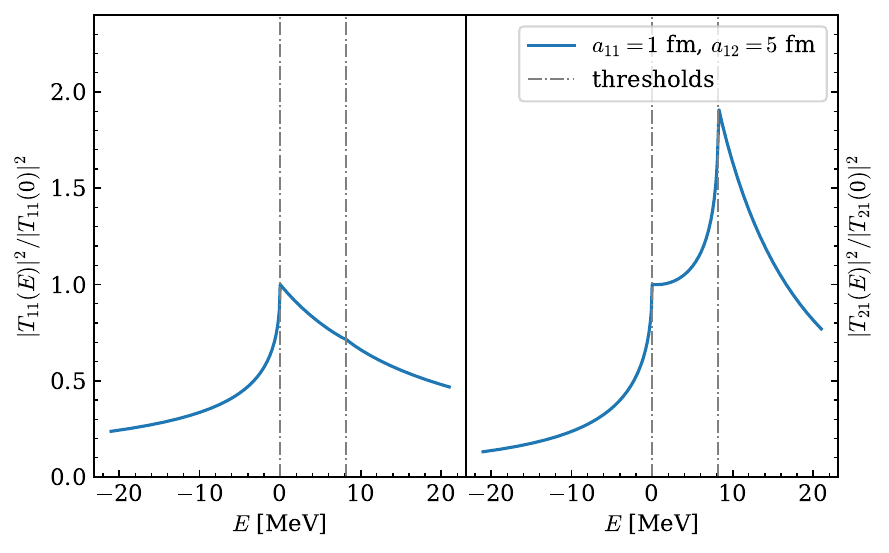} & \SetCell[r=2]{m,0.38\textwidth} \includegraphics[width=\linewidth]{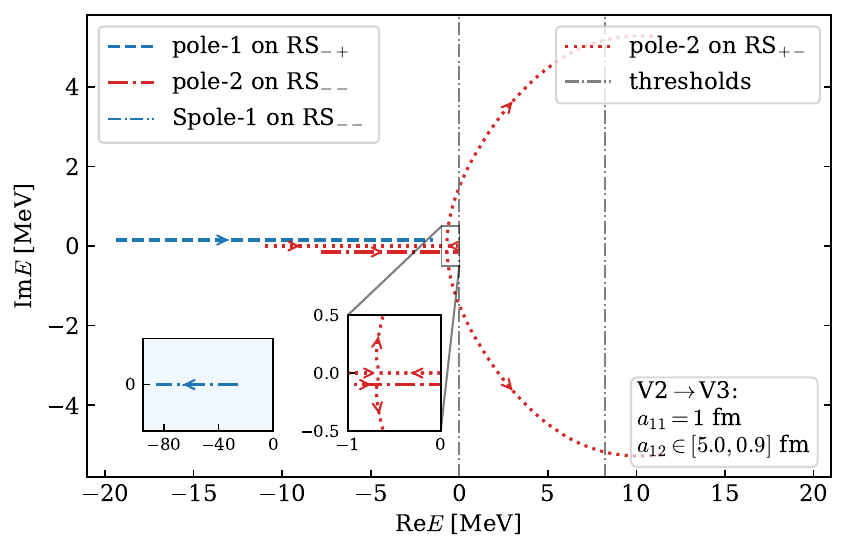}\\ & & & & \\ \cline{1-4} \\
        \SetCell[r=2]{m}V3 & \SetCell[r=2]{m} $(0,1)$ & \SetCell[r=2]{m} $(0,1)$ &
        \SetCell[r=2]{m,0.4\textwidth} \includegraphics[width=\linewidth]{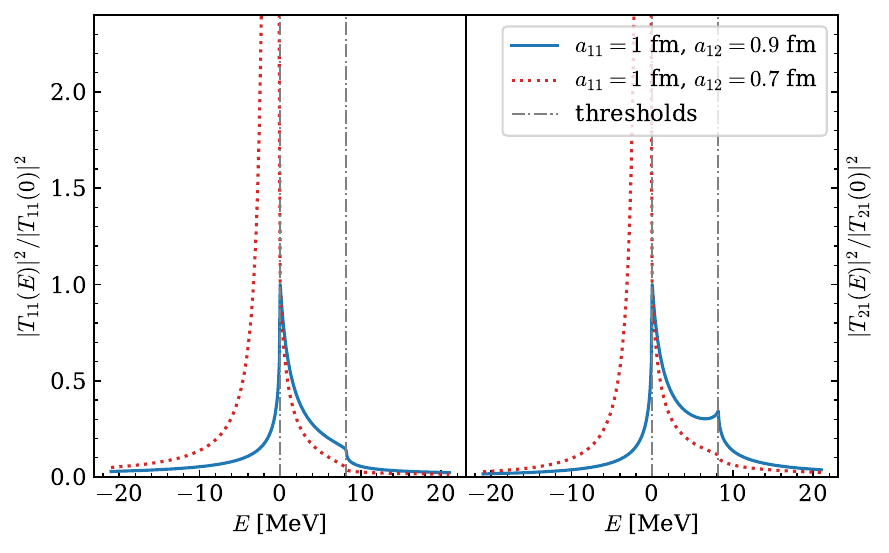} & \SetCell[r=2]{m,0.38\textwidth} \includegraphics[width=\linewidth]{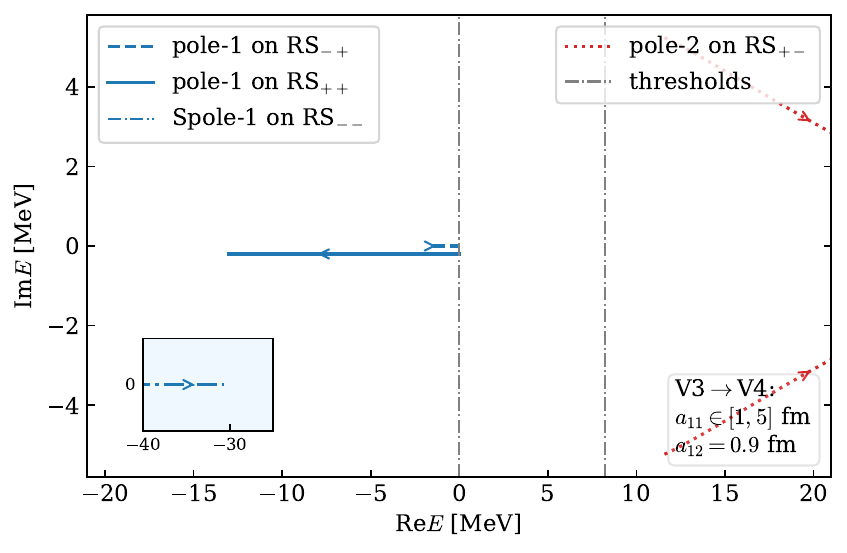}\\ & & & & \\ \cline{1-4} \\
        \SetCell[r=2]{m}V4 & \SetCell[r=2]{m} $\gg 1$ & \SetCell[r=2]{m} $(0,1)$ &
        \SetCell[r=2]{m,0.38\textwidth} \includegraphics[width=\linewidth]{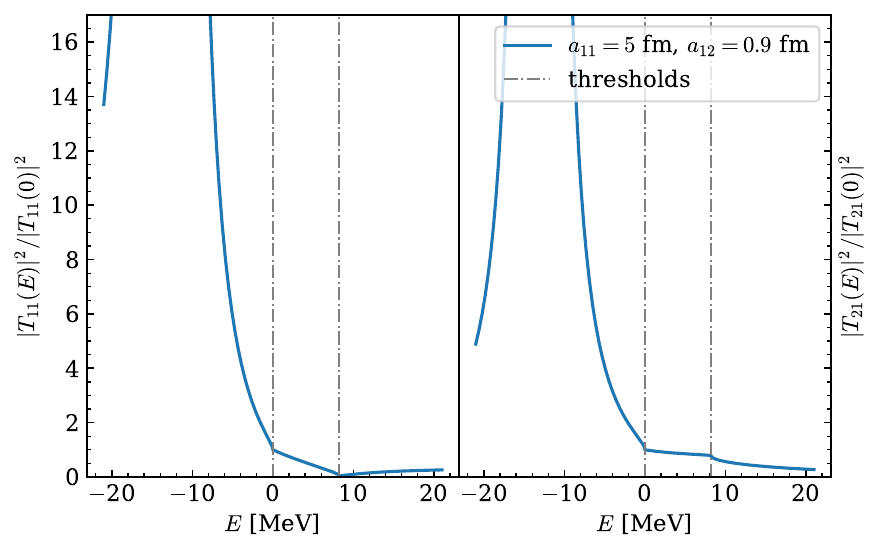} & \SetCell[r=2]{m,0.38\textwidth} \includegraphics[width=\linewidth]{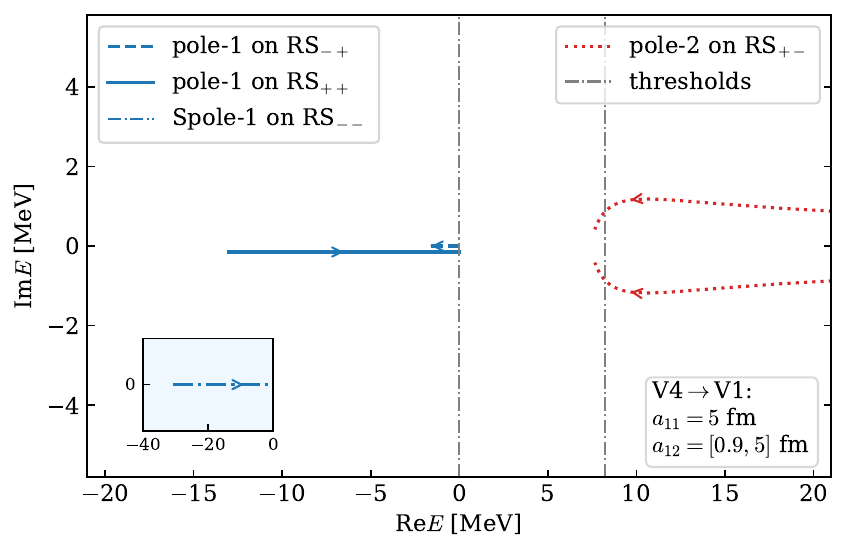}\\ 
        & & & & \\ \hline\hline
    \end{tblr} 
    \label{tab:classification_vs}
\end{table*}

Let us now consider the evolution from the two-pole FP, Eq.~\eqref{eq:2ndFP}, with the variation of $a_{11}$ and $a_{12}$. 
The poles of the amplitudes are given by the solutions of 
\begin{align}
    \left(\frac{1}{a_{11}}-i{r_1} p_1\right)\left(\frac{1}{a_{11}}-i{r_2} p_2\right)-\frac{1}{a_{12}^2} = 0,
    \label{eq:det}
\end{align} 
which implies again the pole duality, {\it i.e.}, the amplitudes with the same $|a_{12}|$ but opposite $a_{11}$ have poles on RS$_{r_{1}\,r_{2}}$ and RS$_{-r_{1}\,-r_{2}}$, respectively, at the same positions.
From
\begin{align}
    \mathbf{T}_{21}(E)=\frac{-2\pi a_{12}^{-1}}{\sqrt{M_1M_2}\mathrm{det}},\quad  \mathbf{T}_{11}(E)=\frac{-2\pi\left({a_{11}^{-1}}-ip_2\right)}{M_1\mathrm{det}},
    \label{eq:T21T11}
\end{align}
with $\mathrm{det}\equiv \left(a_{11}^{-1}-i p_1\right)\left(a_{11}^{-1}-ip_2\right)-a_{12}^{-2}$,
the line shape of $\mathbf{T}_{21}$ is dominated by the poles, while that of $\mathbf{T}_{11}$ is complicated due to a zero at $E=\Delta\left(1-{a_{11}^{-2}\delta^{-2}}\right)$ in addition~\cite{Dong:2020hxe} (see also Ref.~\cite{Sone:2024nfj}).

At the RG FP in Eq.~\eqref{eq:2ndFP}, $a_{11}=a_{12}=\infty$ and $\delta=0$, there are two bound/virtual state poles at $E=0$,
the threshold of both channels. The two poles will separate and be located at $E=0$ and $E=\Delta$, respectively, for $\delta\neq 0$. 
For finite $a_{11}$ and $a_{12}$, using the conformal mapping~\cite{Kato:1965iee,Guo:2016bjq,Yamada:2021azg}, 
\begin{align}
    p_1 = \sqrt{\frac{\mu_{1}\Delta}{2}}\left(\omega+\frac{1}{\omega}\right), 
    \quad p_2 = \sqrt{\frac{\mu_{2}\Delta}{2}}\left(\omega-\frac{1}{\omega}\right),
\end{align} 
the four RSs of the complex $E$ plane can be mapped into the $\omega$ plane. It is easy to find that there are four poles in the $\omega$ plane in total, and thus four poles in all RSs of $E$ plane. 

Then we can classify the near-threshold line shapes according to the values of $a_{11}$ and $|a_{12}|$. Cases with $a_{11}$ negative and positive are labeled by B and V, corresponding to having bound and virtual state poles in the single-channel situation, respectively. The line shapes in these two groups of cases are shown in Tables~\ref{tab:classification_bs} and \ref{tab:classification_vs}. 
One sees that the line shapes can be quite different for different values of $a_{11}$ and $|a_{12}|$, however, the evolution can be understood from the pole trajectories, which are shown in the tables as well.
The tables may be regarded as a dictionary for the near-threshold line shapes and the corresponding pole locations.

Starting from the near-FP situation $a_{11}\delta\ll -1$ and $1/a_{12}=0$, each channel has a bound state just below the corresponding threshold. Without channel coupling, each bound state has two poles on different RSs at the same location. That is, the poles of the channel-1 bound state are on RS$_{++}$ and RS$_{+-}$, while those of the channel-2 bound state are on RS$_{++}$ and RS$_{-+}$.

Case B1 ($a_{11}\delta\ll -1$, $|a_{12}|\delta \gg 1$) is obtained by switching on the channel coupling. Both channel-1 poles are pushed downward on their RSs; the one on RS$_{++}$ (pole-1 in Table~\ref{tab:classification_bs}) is the main pole since it is close to the physical region while the one on RS$_{+-}$ becomes its shadow (Spole-1)~\cite{Eden:1964zz} and has little effect on the physical line shape. The channel-2 poles acquire imaginary parts because of the coupling to the lower channel, and become a complex conjugated pair on RS$_{-+}$ (pole-2), as required by the Schwarz reflection principle.
In this case, the line shape of $|\mathbf{T}_{21}|$ has two sharp peaks below the two thresholds, while that of $|\mathbf{T}_{11}|$ has a dip just below the higher threshold, as a consequence of the zero of $\mathbf{T}_{11}(E)$ in Eq.~\eqref{eq:T21T11}~\cite{Dong:2020hxe}.

Case B2 ($-1<a_{11}\delta <0$, $|a_{12}|\delta \gg 1$) is obtained by increasing $a_{11}$ (corresponding to increasing the single-channel attraction) from Case B1. Both poles move downward. Eventually, the pole-2 conjugated pair move down below the lower threshold on RS$_{-+}$, then one of them moves upward, transiting to RS$_{++}$ at the lower threshold, and the other moves further downward on RS$_{-+}$ and has little effect on the physical line shape. Correspondingly, two peaks appear in the line shapes.

Case B3 ($-1<a_{11}\delta <0$, $|a_{12}|\delta < 1$) is obtained by decreasing $|a_{12}|$ (increasing channel coupling, which introduces an effective attraction in channel-1 and repulsion in channel-2) from Case B2. Pole-1 moves further downward and remotely, while pole-2 moves upward and eventually back to RS$_{-+}$ as a conjugated pair. If pole-2 is above the higher threshold, a sharp threshold cusp at the higher threshold appears (blue solid lines in the line shapes of Case B3 in Table~\ref{tab:classification_bs}); if pole-2 is between the two thresholds, a bump appears in the line shapes (red dotted lines). 

Case B4 ($a_{11}\delta \ll -1$, $|a_{12}|\delta < 1$) is obtained by increasing $|a_{11}|$ from Case B3. Pole-1 moves upward, and pole-2 moves upward far above the higher threshold. It can happen that pole-1 is close to the lower threshold, or all poles are far from the thresholds; in the latter situation, no prominent near-threshold peak appears in the line shapes. The zero of $\mathbf{T}_{11}(E)$ can still give a dip.

The positive $a_{11}$ cases can be analyzed similarly and are listed in Table~\ref{tab:classification_vs}.
One distinction compared to the above cases is that the line shapes are threshold cusps except for Case V4 where pole-1 is dragged by the effective attraction due to the strong channel coupling to RS$_{++}$ and gives a peak below the lower threshold.

Taking the $X(3872)$ case as an example, we show how to apply our classification to get some information about the near threshold resonant states. The $X(3872)$ appears as a pronounced peak in the $J/\psi \pi^+\pi^-$ and $D^0\bar{D}^0\pi^0$ invariant mass distribution~\cite{Belle:2003nnu,LHCb:2020xds,LHCb:2020fvo,BESIII:2023hml} close to the $D^0\bar{D}^{*0}$ threshold and about $8$~MeV below the $D^+D^{*-}$ threshold, which is related to the threshold of $D^0\bar{D}^{*0}$ via isospin symmetry. 
Given that experiments have observed only one prominent peak, we can deduce from Tables~\ref{tab:classification_bs} and \ref{tab:classification_vs} that the $X(3872)$ case must correspond to either Case B4, V3, or V4. These cases are characterized by a pronounced peak appearing just below or at the threshold of channel-1. 
A recent detailed analysis~\cite{Zhang:2024fxy} showed that the $J^{PC}=1^{++}$ $D^0\bar D^{*0}$-$D^+D^{*-}$ coupled-channel system with the $X(3872)$ corresponds to Case V4:
there are two virtual state poles, one near the $D^0\bar D^{*0}$ threshold and the other near the $D^+D^{*-}$ threshold if the channel coupling is switched off. The strong channel coupling pushes the lower pole, corresponding to the $X(3872)$, to become a bound state pole and the higher one, corresponding to the isovector $I^G(J^{PC})=1^-(1^{++})$ $W_{c1}^0$, to above the $D^+D^{*-}$ threshold on RS$_{+-}$. For details, see Appendix~B of Ref.~\cite{Zhang:2024fxy}.\footnote{For precise determinations of the pole positions of the $X(3872)$ and $W_{c1}^0$ from a combined analysis, we refer to Ref.~\cite{Ji:2025hjw}. The charged $W_{c1}^\pm$ as a virtual state  as predicted in Ref.~\cite{Zhang:2024fxy} and has been supported by a recent lattice calculation~\cite{Sadl:2024dbd}.}


In all the cases, if the $|\mathbf{T}_{i1}|$ ($i=1,2$) line shapes drop monotonically below the lower threshold and above the higher threshold, the width of the peaking structure around the thresholds is controlled by the threshold splitting $\Delta$. To get a quantitative understanding of the widths of the peaking structures, we define $E_+$ and $E_-$ as the energies satisfying $|\mathbf{T}_{i1}(E_+)|=|\mathbf{T}_{i1}(\Delta)|/2$ and $|\mathbf{T}_{i1}(E_-)|=|\mathbf{T}_{i1}(0)|/2$, respectively, and the half-threshold-heights width as $\Gamma\equiv E_+-E_-$. Then, we have the following results for $\Gamma$ of $|\mathbf{T}_{i1}|$:
\begin{align}
    \Gamma_{|\mathbf{T}_{21}|\, \text{(V1)}} & = \Delta \left[1+ \mathcal{O}\left(a_{11}^{-2}\delta^{-2}\right)\right], \nonumber \\
    \Gamma_{|\mathbf{T}_{11}|\, \text{(V2)}} & = 4\Delta\left[\frac{1}{(a_{11}\delta)^2}+\frac{1}{a_{11}\delta} + 1+\mathcal{O}\left(a_{11}^2\delta^2\right)\right], \nonumber \\
   \Gamma_{|\mathbf{T}_{21}|\, \text{(V2)}} & = \Delta\left[\frac{7}{4(a_{11}\delta)^2}+\frac{1}{a_{11}\delta} + 3+\mathcal{O}\left(a_{11}^2\delta^2\right)\right],  \nonumber\\
    \Gamma_{|\mathbf{T}_{i1}|\, \text{(B3, V3)}} & = \frac{17}{8}\Delta\left[1+\mathcal{O}\left(|a_{11}|\delta\right)\right],
\end{align}
where the corresponding cases are given in the parentheses.

One sees that the widths of the peaking structures are mainly determined by $\Delta$ for cases V1, and B3, V3. 
For V1, the reason is that the channels are weakly coupled and the poles are in the immediate vicinities of both thresholds, thus the shapes of $|\mathbf{T}_{21}|$ is simply determined by the threshold difference $\Delta$.
While for cases B3 and V3, as can be seen from the ending points of the pole trajectories of the second plots in the rightmost columns of Tables~\ref{tab:classification_bs} and \ref{tab:classification_vs}, or from the beginning points of the pole trajectories of the third plots in the same columns, the lower pole is located below the lower threshold and the higher pole is above the higher threshold.
They are either far away from the threshold (for the lower pole in B3) or located on a remote RS (for the other poles) in the sense that they need to circle around the corresponding thresholds to reach the physical region. Furthermore, the channel coupling is strong ($|a_{12}|\delta<$1). These factors induce the line shapes to be determined by the threshold difference as well.
The width in Case V2 also has strong dependence on the value of $a_{11}$.
For case V2, the channel coupling is small, and the higher pole is located blow the lower threshold due to the value of $a_{11}\delta \in (0,1)$. Consequently the value of $a_{11}$ plays a role together with the threshold difference in forming the $|\mathbf{T}_{ij}|$ line shapes.

\medskip

\section{Summary}

To summarize, the general near-threshold structures in the coupled-channel system have been classified according to the single-channel scattering lengths and the channel coupling strength, with the evolution starting from the RG FPs.
A symmetry-related two-channel system was discussed as an example, and a dictionary for the evolution of line shapes and the corresponding pole locations with the variation of the single-channel scattering lengths and the channel couplings was provided in Tables~\ref{tab:classification_bs} and \ref{tab:classification_vs}.
The results can be used to perceive rough pole locations from line shapes, and thus useful for understanding the complicated line shapes of exotic hadron candidates in the near-threshold region which were observed in recent years and are expected to be observed in the future.

Considering the phase space, the line shapes below the lower threshold considered here can only be observed in the final states of a lower channel. A three-channel discussion is needed for a complete treatment, and the results are similar as reported here if the lowest channel is weakly coupled to the higher two channels~\cite{Zhang:2023zzg}.

\section*{Acknowledgements}
    This work is supported in part by the Chinese Academy of Sciences under Grants No. YSBR-101 and No.~XDB34030000; by the National Key R\&D Program of China under Grant No. 2023YFA1606703; by the National Natural Science Foundation of China (NSFC) under Grants No. 12125507, No. 12361141819 and No. 12047503; and by NSFC and the Deutsche Forschungsgemeinschaft (DFG) through the funds provided to the Sino-German  Collaborative Research Center CRC110 ``Symmetries and the Emergence of Structure in QCD'' (DFG Project-ID 196253076).

\bibliographystyle{elsarticle-num}
\bibliography{ZREFT.bib}

\clearpage
\begin{appendix}
    
    \section{Details of pole evolution with the variation of \texorpdfstring{$a_{ij}$}{aij}}
    \label{supp:details}
    
    In this supplemental material, we give some detailed descriptions of the pole evolution from the renormalization group fixed point (RG FP) with the variation of $a_{11}$ and $a_{12}$, i.e., the single-channel scattering length and the channel coupling strength, shown in Tables~I and II in the main text.
    
    In the main text, the cases with $a_{11}<0$ and $a_{11}>0$ are labeled as Case B and Case V, and are shown in Tables~I and II, respectively. In these two tables, the poles on the RS$_{++}$, RS$_{-+}$, RS$_{--}$ and RS$_{+-}$ are represented by the solid, dashed, dot-dashed, and dotted lines, respectively. The lower pole is called {\color{blue} pole-1} and marked in blue, and the higher pole is called {\color{red} pole-2} and marked in red. The shadow pole of {\color{blue} pole-1} is named as {\color[RGB]{38,157,247} Spole-1} and shown in the bottom-left blue panel. 
    
    \subsection{Case B (\texorpdfstring{$a_{11}<0$}{a11<0})}
    
    \textbf{Case B1} ($a_{11}\delta\ll-1,|a_{12}|\delta\gg1$)
    
    Near the RG FP, there are three poles on different Riemann sheets (RS):
    
    The {\color{blue} pole-1} on RS$_{++}$ just below the threshold of channel-1, represented by the blue solid line.
    
    The {\color{red} pole-2} on RS$_{-+}$ just below the threshold of channel-2, represented by the red dashed line. The {\color{red} pole-2} has imaginary parts and appears as a complex conjugated pair on the complex energy plane.
    
    The {\color[RGB]{38,157,247} Spole-1} (shadow of {\color{blue} pole-1}) on RS$_{+-}$ just below the threshold of channel-1, represented by the blue dotted line.
    
    \textbf{Case B1} ($a_{11}\delta\ll-1,|a_{12}|\delta\gg1$) $\rightarrow$ \textbf{Case B2} ($-1<a_{11}\delta<0,|a_{12}|\delta\gg1$)
    
    Case B2 is obtained by increasing $a_{11}$ from Case B1. As $a_{11}$ increasing, the trajectories of the poles in Case B1 are:
    
    {\color{blue} pole-1} on RS$_{++}$ and {\color[RGB]{38,157,247} Spole-1} on RS$_{+-}$: Move downward and further from the threshold.
    
    {\color{red} pole-2} on RS$_{-+}$: The conjugated pair move downward to the threshold of channel-1. When they move close to the threshold of channel-1, the pair meet at the real $E$ axis below the threshold. Then one of the two poles in the pair keeps on the RS$_{-+}$ and moves downward away from the threshold, and the other pole moves upward to the channel-1 threshold. The pole moving upward transits to RS$_{++}$ when it meets the channel-1 threshold, and then it moves downward on RS$_{++}$, further from the threshold.
    
    \textbf{Case B2} ($-1<a_{11}\delta<0,|a_{12}|\delta\gg1$)
    $\rightarrow$ \textbf{Case B3} ($-1<a_{11}\delta<0,|a_{12}|\delta<1$)
    
    Case B3 is obtained by decreasing $|a_{12}|$ from Case B2. As $|a_{12}|$ decreasing, the trajectories of the poles in Case B2 are:
    
    {\color{blue} pole-1} on RS$_{++}$: Moves downward and further from the threshold.
    
    {\color{red} pole-2}: There is one pole on the RS$_{++}$ and another pole on RS$_{-+}$ below the channel-1 threshold. Both poles move upward to the threshold. The pole on RS$_{++}$ moves to the RS$_{-+}$ when it meets the threshold, then moves downward. Then the two poles on RS$_{-+}$ meet below the channel-1 threshold on the real $E$ axis, and become a complex conjugated pair on the complex energy plane. This pair move upward with $|a_{12}|$ decreasing.
    
    {\color[RGB]{38,157,247} Spole-1} on RS$_{+-}$: Moves upward to the channel-1 threshold.
    
    \textbf{Case B3} ($-1<a_{11}\delta<0,|a_{12}|\delta<1$)
    $\rightarrow$ \textbf{Case B4} ($a_{11}\delta\ll-1,|a_{12}|\delta<1$)
    
    Case B4 is obtained by decreasing $a_{11}$ from Case B3. As $a_{11}$ decreasing, the trajectories of the poles in Case B3 are:
    
    {\color{blue} pole-1} on RS$_{++}$: Moves upward to the channel-1 threshold.
    
    {\color{red} pole-2} on RS$_{-+}$: The complex conjectured pair move upward and further from the thresholds.
    
    {\color[RGB]{38,157,247} Spole-1} on RS$_{+-}$: Moves upward to the channel-1 threshold. It transits to RS$_{--}$ when it meets the channel-1 threshold, then it moves downward.
    
    \textbf{Case B4} ($a_{11}\delta\ll-1,|a_{12}|\delta<1$)
    $\rightarrow$ \textbf{Case B1} ($a_{11}\delta\ll-1,|a_{12}|\delta\gg1$)
    
    Case B1 is obtained by increasing $|a_{12}|$ from Case B4. As $|a_{11}|$ increasing, the trajectories of the poles in Case B4 are:
    
    {\color{blue} pole-1} on RS$_{++}$: Moves upward to the channel-1 threshold.
    
    {\color{red} pole-2} on RS$_{-+}$: The complex conjectured pair move downward and close to the channel-2 threshold.
    
    {\color[RGB]{38,157,247} Spole-1} on RS$_{--}$: Moves upward to the channel-1 threshold. It transits to RS$_{+-}$ when it meets the channel-1 threshold, then it moves downward.
    
    \subsection{Case V (\texorpdfstring{$a_{11}>0$}{a11>0})}
    
    \textbf{Case V1} ($a_{11}\delta\gg 1,|a_{12}|\delta\gg1$)
    
    Near the RG FP, there are three poles on different Riemann sheets (RS):
    
    The {\color{blue} pole-1} on RS$_{-+}$ just below the threshold of channel-1, represented by the blue dashed line.
    
    The {\color{red} pole-2} on RS$_{+-}$ just below the threshold of channel-2, represented by the red dotted line. The {\color{red} pole-2} has imaginary parts and appears as a complex conjugated pair on the complex energy plane.
    
    The {\color[RGB]{38,157,247} Spole-1} (shadow of {\color{blue} pole-1}) on RS$_{--}$ just below the threshold of channel-1, represented by the blue dot-dashed line.
    
    \textbf{Case V1} ($a_{11}\delta\gg 1,|a_{12}|\delta\gg1$) $\rightarrow$ \textbf{Case V2} ($0<a_{11}\delta<1,|a_{12}|\delta\gg1$)
    
    Case V2 is obtained by decreasing $a_{11}$ from Case V1. As $a_{11}$ decreasing, the trajectories of the poles in Case V1 are:
    
    {\color{blue} pole-1} on RS$_{-+}$ and {\color[RGB]{38,157,247} Spole-1} on RS$_{--}$: Move downward and further from the threshold.
    
    {\color{red} pole-2} on RS$_{+-}$: The conjugated pair move downward to the threshold of channel-1. When they move close to the threshold of channel-1, the pair meet at the real $E$ axis below the threshold. Then one of the two poles in the pair keeps on the RS$_{+-}$ and moves downward away from the threshold, and the other pole moves upward to the channel-1 threshold. The pole moving upward transits to RS$_{--}$ when it meets the channel-1 threshold, and then it moves downward on RS$_{--}$, further from the threshold.
    
    \textbf{Case V2} ($0<a_{11}\delta<1,|a_{12}|\delta\gg1$)
    $\rightarrow$ \textbf{Case V3} ($0<a_{11}\delta<1,|a_{12}|\delta<1$)
    
    Case V3 is obtained by decreasing $|a_{12}|$ from Case V2. As $|a_{12}|$ decreasing, the trajectories of the poles in Case V2 are:
    
    {\color{blue} pole-1} on RS$_{-+}$: Moves upward to the channel-1 threshold.
    
    {\color{red} pole-2}: There is one pole on the RS$_{--}$ and another pole on RS$_{+-}$ below the channel-1 threshold. Both poles move upward to the threshold. The pole on RS$_{--}$ moves to the RS$_{+-}$ when it meets the threshold, then moves downward. Then the two poles on RS$_{+-}$ meet below the channel-1 threshold on the real $E$ axis, and become a complex conjugated pair on the complex energy plane. This pair move upward with $|a_{12}|$ decreasing.
    
    {\color[RGB]{38,157,247} Spole-1} on RS$_{--}$: Moves downward and further from the threshold.
    
    \textbf{Case V3} ($0<a_{11}\delta<1,|a_{12}|\delta<1$)
    $\rightarrow$ \textbf{Case V4} ($a_{11}\delta\gg 1,|a_{12}|\delta<1$)
    
    Case V4 is obtained by increasing $a_{11}$ from Case V3. As $a_{11}$ increasing, the trajectories of the poles in Case V3 are:
    
    {\color{blue} pole-1} on RS$_{-+}$: Moves upward to the channel-1 threshold. It transits to RS$_{++}$ when it meets the channel-1 threshold, then it moves downward.
    
    {\color{red} pole-2} on RS$_{+-}$: The complex conjectured pair move upward and further from the thresholds.
    
    {\color[RGB]{38,157,247} Spole-1} on RS$_{--}$: Moves upward to the channel-1 threshold.
    
    \textbf{Case V4} ($a_{11}\delta\gg 1,|a_{12}|\delta<1$)
    $\rightarrow$ \textbf{Case V1} ($a_{11}\delta\gg 1,|a_{12}|\delta\gg1$)
    
    Case V1 is obtained by increasing $|a_{12}|$ from Case V4. As $|a_{12}|$ increasing, the trajectories of the poles in Case V4 are:
    
    {\color{blue} pole-1} on RS$_{++}$: Moves upward to the channel-1 threshold. It transits to RS$_{-+}$ when it meets the channel-1 threshold, then it moves downward.
    
    {\color{red} pole-2} on RS$_{+-}$: The complex conjectured pair move downward and close to the channel-2 threshold.
    
    {\color[RGB]{38,157,247} Spole-1} on RS$_{--}$: Moves upward to the channel-1 threshold.

    \end{appendix}
    
\end{document}